%% file: main.tex
\documentclass{acm_sen_article}
\usepackage[utf8]{inputenc}
\usepackage{pgfgantt}
\usepackage{enumitem}
\usepackage{xparse}
\usepackage{todonotes}
\usepackage{flushend}
\usepackage{caption}
\usepackage{subcaption}
\usepackage{authblk}

\newcommand\textganttbar[5][]{%
    \ganttbar[#1,bar/.append style={alias=tmp}]{#2}{#4}{#5}
    \path 
    let
    \p1=(tmp.west),\p2=(tmp.east),
    \n1={\x2-\x1},\n2={width("#3")},
    \n3={ifthenelse(\n1>\n2,90,270)}
    in
    node [anchor=\n3,font=\footnotesize] at (tmp.north) {#3};
}

\NewDocumentCommand\textganttbarB{O{} O{} mmmm}{%
    \ganttbar[#1,bar/.append style={alias=tmp}]{#3}{#5}{#6}
    \node [font=\footnotesize,at={(tmp)},#2]  {#4};
}

\tikzset{
  above bar/.style={
    at={(tmp.north)},anchor=south
    },
  below bar/.style={
    at={(tmp.south)},anchor=north
    }
}

\title{Methodological Issues in Observational Studies}
\subtitle{Early Stage}

\author{Nyyti Saarimäki \\
Tampere University, 
Tampere, Finland \\
nyyti.saarimaki@tuni.fi
}

\begin{document}

\maketitle
\begin{abstract}
\textbf{Background.} Starting from the 1960s, practitioners and researchers have looked for ways to empirically investigate new technologies such as inspecting the effectiveness of new methods, tools, or practices. With this purpose, the empirical software engineering domain started to identify different empirical methods, borrowing them from various domains such as medicine, biology, and psychology. Nowadays, a variety of empirical methods are commonly applied in software engineering, ranging from controlled and quasi-controlled experiments to case studies, from systematic literature reviews to the newly introduced multivocal literature reviews. However, to date, the only available method for proving any cause-effect relationship are controlled experiments.
\newline\textbf{Objectives.}  The goal of the thesis is introducing new methodologies for studying causality in empirical software engineering. 
\newline\textbf{Methods.} Other fields use observational studies for proving causality. They allow observing the effect of a risk factor and testing this without trying to change who is or is not exposed to it. As an example, with an observational study it is possible to observe the effect of pollution on the growth of a forest or the effect of different factors on development productivity without the need of waiting years for the forest to grow or exposing developers to a specific treatment. 
\newline\textbf{Conclusion.} In this thesis, we aim at defining a methodology for applying observational studies in empirical software engineering, providing guidelines on how to conduct such studies, how to analyze the data, and how to report the studies themselves. 
\end{abstract}


\input{1_introduction.tex}
\input{2_related_works.tex}

\input{3_proposed_approach.tex}

\input{4_current_status.tex}


\section{Expected Contribution}

Currently, researchers in empirical software engineering have access to large quantities of data. However, many studies concentrate on correlational findings, as there is no commonly accepted research methodology for performing causal studies without running controlled experiments. 

The main contribution of this thesis will be the introduction of a new set of methodologies in the domain of empirical software engineering as an extension of existing methodologies applied in other domains such as medicine, biology, and psychology. 
The methodologies will be proposed together with user guidelines on how to perform such studies and how to report them. 
The second main contribution of this work will be the identification of a set of validated data analysis techniques for handling dependent data, which, besides being used in the methodologies proposed, could also be beneficial for researchers applying different methodologies and needing to analyze dependent data. 


\bibliographystyle{IEEEtran}
\bibliography{references}

\end{document}

%% file: 1_introduction.tex
\section{Introduction}

Software engineering is a relatively new field of research compared to other engineering disciplines such as mechanical engineering or civil engineering. It was formed in the 1960s when developers realized that understanding the code is not enough when creating a piece of software.
From that time on, software engineering research started to focus on different aspects, including the definition of new processes (e.g., the spiral model ~\cite{Boehm1988}. More recently, the focus has been on topics such as agile~\cite{beck2001agile} and lean models~\cite{Poppendieck2007}), testing approaches (e.g., test-driven development~\cite{Beck2002TDD}),  development tools such as IDEs, specific techniques such as reading techniques or Fault Trees~\cite{Boyd1991}, and many others. 
Nowadays, software engineering covers all aspects related to engineering software, and covers the whole lifecycle of a program.

The introduction of all these new technologies has created the need to validate them. Starting from the 1980s, different groups have proposed the application of empirical methods already adopted in different disciplines, such as medicine, biology, and psychology, to the newly proposed software technologies. This led to the birth of “empirical software engineering”.


The research methodologies adopted from other domains include controlled and quasi-experiments~\cite{Basili1986}\cite{Jedlitschka08}, case studies~\cite{Runeson2009}, systematic literature reviews~\cite{Kitchenham07}, and most recently, multivocal literature reviews~\cite{GAROUSI2019}. 

While these studies broaden our understanding of the studied aspects, it has not been proven that the results of such studies can be generalized. Thus, the root causes of the different problems remain unknown. In software engineering, it is generally believed that causality cannot be proven without controlled experiments. However, researchers in the medical domain have adopted ''observational studies'' to investigate causality retrospectively, without the need to run controlled experiments. This raises the question whether such methodologies could be applied in empirical software engineering as well.

The application of observational studies could be highly beneficial in several software engineering fields, from mining software repositories to effort estimation studies. As an example, software engineering could benefit from approaches used in the epidemiological domain. Epidemiology  studies the distributions of demographic, geographic, and temporal factors in order to understand the causal relationships between exposures and outcomes. Instead of focusing on individuals, it studies populations in order to provide generalizable results that can be applied on a large scale. Several different kinds of studies are commonly used in this field, from observational to experimental studies. These have been used, for example, to study environmental exposures, infectious diseases, and natural disasters. 

Last year, the first set of observational studies was published, primarily based on cohort methods~\cite{Fucci:2018Cohort}\cite{Janzen}. However, the two studies applied different methodologies, mainly because of the lack of clear guidelines. They also used traditional analysis techniques commonly adopted in case studies and experiments, instead of the techniques recommended for observational studies~\cite{Harbord}. 

The main goal of this thesis is to determine how different observational studies can be applied in software engineering.

In this work, we will attempt to answer the following research questions (RQs)
\begin{itemize}[itemsep=1pt,topsep=-5pt]
    \item[\textbf{RQ1}.] Which type of observational studies can be applied in empirical software engineering? 
    \item[\textbf{RQ2}.] Which analysis techniques should be applied in the different observational studies? 
    \item[\textbf{RQ3}.] How to report the different observational studies?
\end{itemize}
\newpage

%% file: 2_related_works.tex
\section{Background and Related Work}


Different study methodologies have been proposed in the field of empirical software engineering. The most common ones are:  
\begin{description}[itemsep=1pt,topsep=-5pt]
    \item[Controlled (and quasi-) experiments]~\cite{Basili1986}\cite{Jedlitschka08}, which originated from medicine and are used when researchers want to control the behavior of different factors. Traditionally, there is one group that gets a treatment and a control group that is not treated. In software engineering, an experiment can be either human- or technology-oriented.
    
    \item[Case studies]~\cite{Runeson2009} originated from clinical medicine, from where they have spread to several fields. In case studies, the researcher selects a specific case and uses qualitative and quantitative methods to collect data from it. In software engineering, the case can be, for example, a tool or a process. Case studies are relatively common in empirical software engineering as it is relatively easy to study one project but it takes much more effort to investigate several cases. 

    \item[Systematic literature reviews]~\cite{Kitchenham07} are a type of survey originating from medicine. First, the researcher forms research questions and then answers them based on the research conducted. Essentially, the goal is to provide a synthesis of the results obtained in previous studies in order to provide an overview of a phenomenon.
    
    \item[Multivocal literature reviews]~\cite{GAROUSI2019} are one of the most recent additions to the field of empirical software engineering. They are a type of systematic literature review developed in psychology that support the inclusion of the gray literature. This provides a broader view on the selected research questions.
\end{description}

\subsection{Observational Studies}

Observational studies are widely used in medicine. As shown in Table~\ref{tab:evidence_levels}, traditionally they are considered to provide high level of evidence, bettered only by a high quality randomized controlled trial. Thus observational studies are especially important in cases where controlled experiments are not feasible. For example, when investigating new techniques in plastic surgery, controlled experiments are not always indicated or ethical to conduct. Research in medicine proves that well-designed observational studies can provide similar results as controlled experiments. 

\begin{table*}[]
    \centering
    \begin{tabular}{c|p{15cm}}
        Level of Evidence & Qualifying Studies \\
        \hline
        I & High-quality, multicenter or single-center, randomized controlled trial with adequate power; or systematic review of such studies \\
        II & Lesser quality, randomized controlled trial; prospective cohort study; or systematic review of such studies \\
        III & Retrospective comparative study; case-control study; or systematic review of such studies \\
        IV & Case Series \\
        V & Expert opinion; case report or clinical example; or evidence based on physiology, bench research, or “first principles” \\ \hline 
    \end{tabular}
    \caption{Classification of different research methodologies based on the level of evidence they provide~\cite{chung2009introducing}.}
    \label{tab:evidence_levels}
\end{table*}

Results from observational studies in medicine are often criticized~\cite{song2010observational} for unpredictable confounding factors. However, recent work has shown comparable results for observational studies and controlled experiments~\cite{benson2000comparison}\cite{concato2000randomized}. 
Observational studies can be considered analytic study designs and can be further sub-classified as observational or experimental study designs (Figure~\ref{fig:empirical_studies_medicine}).

The main characteristic differentiating observational and experimental study designs is that in experimental studies, the groups are identified based on the  presence or absence of an intervention. In observational studies, there are no interventions but simply “observations” and assessment of the strength of the relationship between an exposure and a disease variable~\cite{merril2006introduction}. 

Another strength of the observational studies is that the techniques can have different time lines regarding the data. Figure~\ref{fig:observation_timeline} presents how various research methodologies differ from each other in terms of the time line. Cross-sectional studies inspect only the present moment while prospective cohort studies can continue decades into the future. In contrast, retrospective cohort studies and case-control studies are conducted using past events.

The three types of observational studies include:

\subsubsection{Cohort Studies}
Cohort studies~\cite{song2010observational}\cite{grimes2002cohort} originate from medicine and are used to understand how exposure to something affects the development of the outcome. The word cohort originally refers to an ancient Roman military unit, but nowadays in this context it means a “group of people with defined characteristics who are followed up to determine incidence of, or mortality from, some specific disease, all causes of death, or some other outcome.”~\cite{morabia2004history}. 

In cohort studies, the researchers first develop a hypothesis on what exposures might cause the outcome they want to investigate. An example of an exposure-outcome pair in medicine could be smoking and lung cancer; an example in software engineering could be code smells and bugs. The definition of the studied exposure should not leave room for interpretation; for instance, it should be clear whether people who smoke 15 cigarettes a day are considered similar to people who smoke only occasionally or not.  After defining the exposures, the researchers gather two or more groups. One group consists of subjects who are not exposed, while the subjects in the other group(s) are exposed. The researchers follow the subjects of both groups and observe whether they develop the researched outcome. The time frame of the study depends, but they can take from months to decades.

A cohort study can be done prospectively or retrospectively. If the study is done from the present time to the future, it is a prospective study. In a retrospective, however, the data has already been collected and the researcher examines past data. Regardless of when the data is collected, it is always analyzed from exposure to outcome.

Like all methodologies, cohort studies have strengths and weaknesses. A major strength in cohort studies is that the temporal relationships are clear, making it possible to understand causality. The study design also allows studying multiple outcomes at once, and they work well for rare exposures. As an additional bonus, the method allows calculating confidence intervals from the results. However, selection bias is an important issue with this method. This is described in more detail in Section~\ref{sec:biases}. Cohort studies are also sensitive to lost follow-ups and cannot address rare outcomes.

\subsubsection{Case-Control Studies}
\label{sec:case-control-studies}

Case-control studies~\cite{song2010observational}\cite{schulz2002case} were first developed for studying disease etiology and are now widely used in the biomedical field. Case-control studies aim to retrospectively identify factors that contributed to a certain outcome. 

The first step is to select an outcome that is being studied; for example, getting breast cancer or having high fault-proneness. Then two groups are gathered: a group with the selected outcome (cases) and a group whose members do not have the outcome (controls). After the subjects have been selected, researchers retrospectively collect data from them about their exposures. This can be done by conducting interviews or by looking at existing records. The last step consists of comparing the groups. 

Case-control studies are well suited to studying rare outcomes or long latency periods as the outcome is known at the beginning of the study. Compared to cohort studies, they require fewer subjects and are relatively inexpensive. They can also provide an odd-ratio which is a good estimate of the relative risk. However, if the exposure is infrequent, case-control studies become inefficient as it is hard to find any kind of subjects for the study. A major concern with case-studies is that is that if are not done properly, they can suffer from biases stemming from several sources. These are discussed more in detail in Section~\ref{sec:biases}

\subsubsection{Cross-Sectional Studies}
Cross-sectional studies~\cite{levin2006study} can be considered as a snapshot of a population at a given time. Using this methodology, data is collected simultaneously about the exposure and the outcome from all subjects. The subjects are divided into four groups: outcome and exposed, outcome and not exposed, no outcome and exposed, and no outcome and not exposed.

The data collected using this methodology does not contain temporal information and thus cannot be used to prove causality as there is no information about when events took place. Instead, the methodology is used for describing a population and finding the prevalence of outcome and exposure.

\subsubsection{Biases in observational studies}
\label{sec:biases}

As stated before, observational studies are sensitive to several kinds of biases. The described methodologies select groups, gather information about them and finally compare the groups. This is why the selection of the case group and the control group needs to be done with great care in order to avoid selection bias. 

Usually, the case group consists of a subgroup of the population that has the outcome. The inclusion and exclusion criteria need to be clearly defined before selecting the group. For example, having new and old cases in the same group could bias the study, as the same treatments might not have been available earlier. It is also important to be aware of where the cases are selected. If the cases are selected from a subgroup, such as a city, they might not represent the whole population.

Selecting the control group can be even harder than selecting the cases, as researchers have to take into account all potential biases. The control group should be gathered from the same subgroup from which the cases were selected; i.e., if the control group had had the outcome, they would have been chosen as the case group. Second, controls should be chosen independent of the exposure. An example could be a study investigating how the number of sex partners affects the chance of getting AIDS. Using patients from of a STD clinic as controls would bias the results as one could argue such persons have more sex partners than an average person.

As information is collected from the subjects, information bias may occur. It should be paid special attention as it cannot be eliminated with data analysis techniques and it might stem from several different sources. The first source of bias occurs in interviews. The subjects in one group may remember exposures better than the subjects in the other group(s). This is called recall bias and can severely distort the results. The second source of bias is the person collecting the data. All subjects should be treated equally, which is why the data gatherer should not know the main hypothesis of the study or the status of the subject.


\begin{figure*}[t]
\centering
\begin{minipage}{.48\textwidth}
  \centering
  \includegraphics[width=0.95\columnwidth, height=6.3cm]{Figures/methodologies.pdf}
    \captionof{figure}{Time frame in which the described observational studies are conducted~\cite{song2010observational}.}
    \label{fig:observation_timeline}
\end{minipage}%
\hfill
\begin{minipage}{.48\textwidth}
  \centering
    \includegraphics[width=0.95\columnwidth, height=6.3cm]{Figures/empirical_studies.pdf}
    \captionof{figure}{The hierarchy of research methodologies used in medicine.}
    \label{fig:empirical_studies_medicine}
\end{minipage}
\end{figure*}

%% file: 3_proposed_approach.tex
\section{The Proposed Approach }

The PhD plan consists of four main steps: 
\begin{description}[itemsep=1pt,topsep=-5pt]
    \item[Step 1]: Analysis of the literature
    \begin{description}
        \item[Step 1.1] Analysis of the literature on observational studies
        \item[Step 1.2] Analysis of the literature on data analysis in SW
    \end{description}

    \item[Step 2]: Identification of methodologies for applying different observational studies in empirical software engineering 
    \item[Step 3]: Validation of the methodologies   
    \item[Step 4]: Proposal of guidelines for conducting observational studies in empirical software engineering
\end{description}

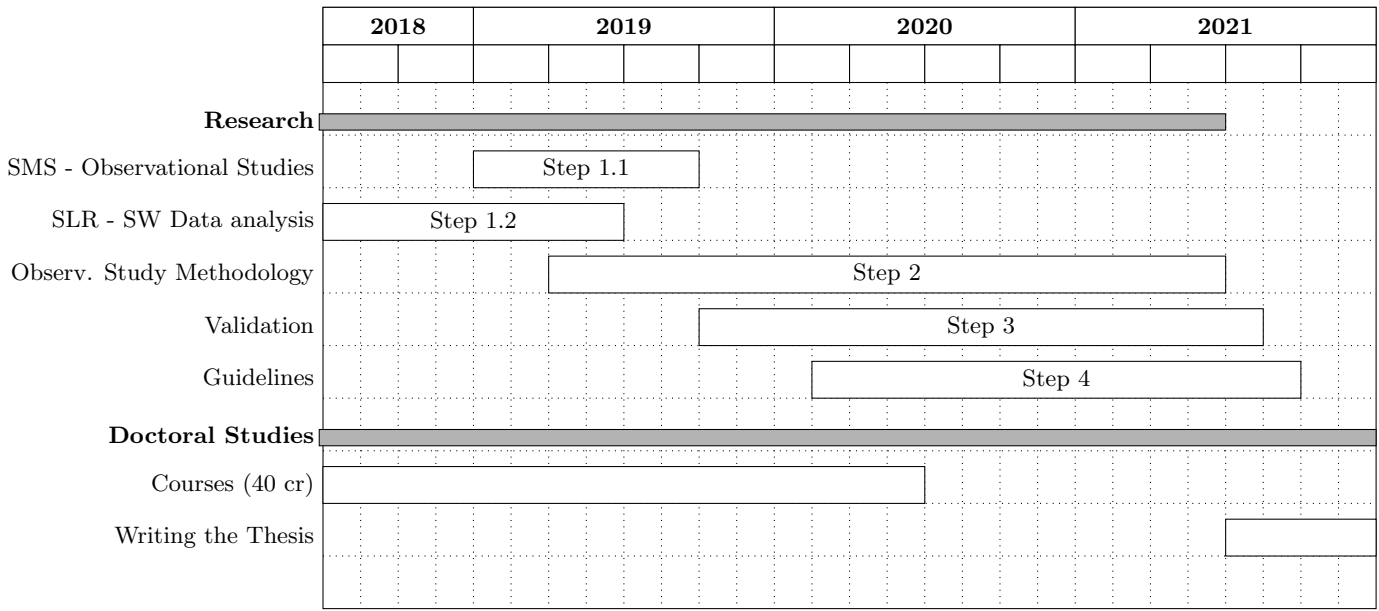
\begin{figure*}[!ht]
    \centering
    \input{timeline.tex}
    \caption{Timeline of the planned doctoral studies}
    \label{fig:timeline}
\end{figure*}

The suitable observational study methodologies for empirical software engineering (RQ1) are defined in steps 1, 2, and 3. The second RQ, concerning the applicability of analysis techniques, is answered in steps 2 and 3. Finally, the last research question (RQ3) is about the reporting, is answered in step 3. The results of the RQs will be combined as a guideline on conducting observational studies in empirical software engineering. The timeline for the PhD is presented in Figure~\ref{fig:timeline}. 

\subsection{Research Methodology}


The four steps will be implemented with different research methods. Each of the steps will be described in more detail below.

\subsubsection{Step 1: Analysis of the literature}

We will first focus on the analysis of the literature on observational studies, classifying them into different groups and understanding how they are applied in different contexts. This step will include the analysis of existing literature in different domains as well as the analysis of software engineering literature to check whether any researcher has ever adopted similar methodologies. 

The latter investigation will include the analysis of the data analysis methodologies used in recent empirical software engineering publications. The goal is to investigate the used research methodologies and types of data as well as how it is analyzed. Especially, it is interesting to know  whether some of the existing studies already adopted similar analysis methodologies. Our hypothesis is that some research done, for example in software repository mining, has already adopted data analysis techniques similar to those required in observational studies. The used statistical analysis techniques have been studied by de Oliveira Neto et al.\cite{DEOLIVEIRANETO2019246}. They confirmed that more attention should be paid to reporting the results from statistical analysis.

\subsubsection{Step 2: Identification of a methodology for applying different observational studies in software engineering} 

In this step, we will propose a set of methodologies that may be suitable for empirical software engineering studies. 
This step will be performed in several rounds. In each round, a different type of observational study will be considered, including cross-sectional, prospective and retrospective cohort, and case-control studies. 

For each type of study, we will first propose an extension of the methodology to be applied. Then we will investigate appropriate data analysis techniques. We assume that the literature review step reveals a need for identifying appropriate data analysis techniques in software engineering. 
Starting with the data analysis techniques that should be adopted in observational studies, we will investigate the most suitable techniques for dealing with dependent data in software repositories. For example, commits of a project are dependent on each other and form a time series and such data could be analyzed using Markov chains and/or time series techniques. 

The different data analysis techniques will first be tested by replicating existing studies. Then we will replicate the same studies again, adopting the observational methodology under investigation (e.g., replicating the study with the same goals, but using an epidemiological method instead of the one adopted in the respective paper).

\subsubsection{Step 3: Validation of the methodologies}

    In this step, we aim at validating the identified methodologies. The goal is to conduct observational studies explaining the cause-effect relationships using the identified methodologies. 
    
    In this step, we will apply the methods identified in Step 2 on actual research data by replicating existing studies. 

    The replicated studies will be selected based on the SLR done in Step 1, where we identified the analysis methodology used in software maintenance and evolution studies. 
    The studies to be replicated must provide the complete dataset and clearly report the data analysis techniques. 
    
    
    Once the replication is completed using the identified techniques, we will compare the results with the original results. We are especially interested in understanding whether the results are the same or whether the observational study methods are able to provide better insights. 
    



\subsubsection{Step 4: Proposal of guidelines for conducting observational studies in software engineering}
Guidelines for conducting studies are widely accepted in software engineering~\cite{Jedlitschka08},~\cite{Runeson2009},~\cite{Kitchenham07},~\cite{GAROUSI2019}. 

Based on the results obtained, we will propose a set of guidelines for conducting and reporting the different types of observational studies. These guidelines will help researchers conducting studies which can prove causality. For example, case studies cannot achieve this and thus guidelines for such studies are not enough.

The guidelines will be based both on existing ones already proposed in medicine~\cite{Gallo2011} and those adopted in software engineering~\cite{Jedlitschka08},~\cite{Runeson2009},~\cite{Kitchenham07},~\cite{GAROUSI2019}. The proposed guidelines will be assessed through peer-review, consulting experts both in software engineering and other domains, and discussing about the replicated results obtained using the guidelines.



\subsection{Threats to Validity}
This work is subject to different threats to validity. First of all, there might not be suitable studies for replication. This might be because of the data used in the study is not applicable, enough data is not collected or because the data is not publicly available. Fortunately, several studies~\cite{Saarimaki2019},~\cite{LenarduzziQCM2017},~\cite{LenarduzziMaltesque2019},~\cite{Lomio2019},~\cite{Janes2017} have used our Technical Debt Dataset. Thus, there is a good possibility we will be able to replicate at least some of them.

As for the different type of observational studies, 
our preliminary investigation, and two  works published in 2018~\cite{Janzen} and 2019~\cite{Fucci:2018Cohort} show at least the possibility of applying cohort studies, while we found several similarities with 
case-control. However, at the moment we are not sure about the applicability of cross-sectional surveys.  

It is possible that the results of the application of observational studies will not change the results of studies conducted with other methods. The same issue could happen for the application of data analysis techniques for dependent data.  
However, the application of proper formal methodologies and more correct data analysis techniques will definitely improve the quality of future work and will enable researchers to consider new types of empirical studies in their research.

%% file: timeline.tex
\begin{ganttchart}[
     y unit title=0.5cm,
     y unit chart=0.7cm,
     vgrid,hgrid,
     title height=1,
     title label font=\bfseries\footnotesize,
     bar height=0.7,
     group/.style={draw=black, fill=black!30},
     group right shift=0,
     group top shift=0.7,
     group height=.1,
     group peaks width={0.2}]{1}{28}
    \gantttitle[]{2018}{4}                 
    \gantttitle[]{2019}{8} 
    \gantttitle[]{2020}{8}                 
    \gantttitle[]{2021}{8} \\ 
    \gantttitle{}{2}
    \gantttitle{}{2}
    \gantttitle{}{2}                      
    \gantttitle{}{2}
    \gantttitle{}{2}
    \gantttitle{}{2}
    \gantttitle{}{2}
    \gantttitle{}{2}
    \gantttitle{}{2} 
    \gantttitle{}{2}
    \gantttitle{}{2}
    \gantttitle{}{2}
    \gantttitle{}{2} 
    \gantttitle{}{2}\\

    

    \ganttgroup[inline=false]{Research}{1}{24} \\ 
    
    \textganttbar[]{SMS - Observational Studies}{Step 1.1}{5}{10} \\
    \textganttbar[]{SLR - SW Data analysis}{Step 1.2}{1}{8} \\
    
    \textganttbar[]{Observ. Study Methodology}{Step 2}{7}{24} \\
    
    \textganttbar[]{Validation}{Step 3}{11}{25} \\
    
    \textganttbar[]{Guidelines}{Step 4}{14}{26}\\

    
    \ganttgroup[inline=false]{Doctoral Studies}{1}{28} \\ 
    
    \ganttbar[inline=false]{Courses (40 cr)}{1}{16}\\
    \ganttbar[inline=false]{Writing the Thesis}{25}{28}\\

\end{ganttchart}

%% file: 4_current_status.tex
\section{Current Status}
We started this PhD in July 2018. We have already investigated Step 1.2, performing a systematic literature review on the methods and analysis techniques adopted in software maintenance and evolution models, and are currently performing a mapping study of the different observational studies. 

We have started the investigation of RQ3, investigating and replicating different empirical studies. We first collected a large dataset, analyzing Java projects from the Apache Software Foundation. 
The data was mined from ASF project repositories and their respective issue trackers. The code quality of the commits was analyzed using SonarQube~\footnote{SonarQube. http://www.sonarqube.org} and Ptidej\footnote{http://www.ptidej.net/}, while fault-inducing and fault-fixing commits were identified using the SZZ algorithm~\cite{SZZ}. In addition to the quality and fault data, the dataset contains information about the commits themselves, such as code complexity, number of lines of code, and date.  
The analysis of the first step of the observational study (composition and diffusion of ''issues'') in the dataset has been recently published~\cite{SaarimakiTechDebt2019}~\cite{LenarduzziPromise2019}. This dataset will be used as a baseline for the next part of the research, together with different datasets adopted in existing studies.


As for the identification of the data analysis technique, we are currently collaborating with different partners
on the identification of appropriate analysis techniques for dependent data in software engineering.
After determining a better way for analyzing commit data, we will investigate how to apply the epidemiological study methodology to replicate a selected set of studies. 


\newpage